\documentclass[11pt, draftclsnofoot, onecolumn]{IEEEtran}

\usepackage{cite,graphicx,amsmath,amssymb}
\usepackage{subfigure}
\usepackage{fancyhdr}
\usepackage{mdwmath}
\usepackage{mdwtab}
\usepackage{balance}
\usepackage{xcolor}
\usepackage{bm}
\usepackage{amsthm}
\usepackage{algorithm}
\usepackage{algorithmic}
\usepackage{multirow}
\usepackage{flafter}

\begin{document}
\title{Low-Power Wide-Area Networks for Sustainable IoT}

\author{
\IEEEauthorblockN{Zhijin~Qin,~
Frank Y. Li,~
Geoffrey Ye Li,~
Julie A. McCann,~
and Qiang Ni
 }
 \thanks{Zhijin Qin is with Queen Mary University of London, London, UK, E1 4NS, email: z.qin@qmul.ac.uk.}
\thanks{Frank Y. Li is with University of Agder, 4898 Grimstad, Norway, email: frank.li@uia.no.}
\thanks{Geoffrey Y. Li is with Georgia Institute of Technology, Atlanta, GA, USA, 30332-0250, email: liye@ece.gatech.edu.}
\thanks{Julie A. McCann is with Imperial College London, London, UK, SW7 2AZ email: jamm@imperial.ac.uk.}
 \thanks{Qiang Ni is with Lancaster University, Lancaster, UK, LA1 4YW, email: q.ni@lancaster.ac.uk.}
}

\maketitle

\begin{abstract}
Low-power wide-area (LPWA) networks are attracting extensive attention  because of their abilities to offer low-cost and massive connectivity to  Internet of Things (IoT) devices distributed over wide geographical areas. This article provides a brief overview on the existing LPWA technologies  and useful insights to aid the large-scale deployment of LPWA networks. Particularly, we first review the currently competing candidates of LPWA networks, such as narrowband IoT (NB-IoT) and long range (LoRa), in terms of technical fundamentals and large-scale deployment potential. Then we present two implementation examples on LPWA networks. By analyzing the field-test results, we identify several challenges that prevent LPWA technologies moving from the theory to wide-spread practice.
\end{abstract}


\section{A Glance at Sustainable IoT}
It has been predicted that the number of industrially connected wireless sensing, tracking, and control devices will approach half a billion in 2025. Such Internet of Things (IoT) networks shall be able to transparently and seamlessly incorporate a massive number of heterogeneous end devices, while providing open access to selected subsets of data for digital services~\cite{Pallavi:2018}. Many of these services allow the user to better understand the system  under focus, i.e., smart city, smart farm, and smart assembly line, and therefore make them greener, more eco-friendly, efficient, and cost effective.

\begin{figure}[t!]
    \begin{center}
        \includegraphics[width=5in]{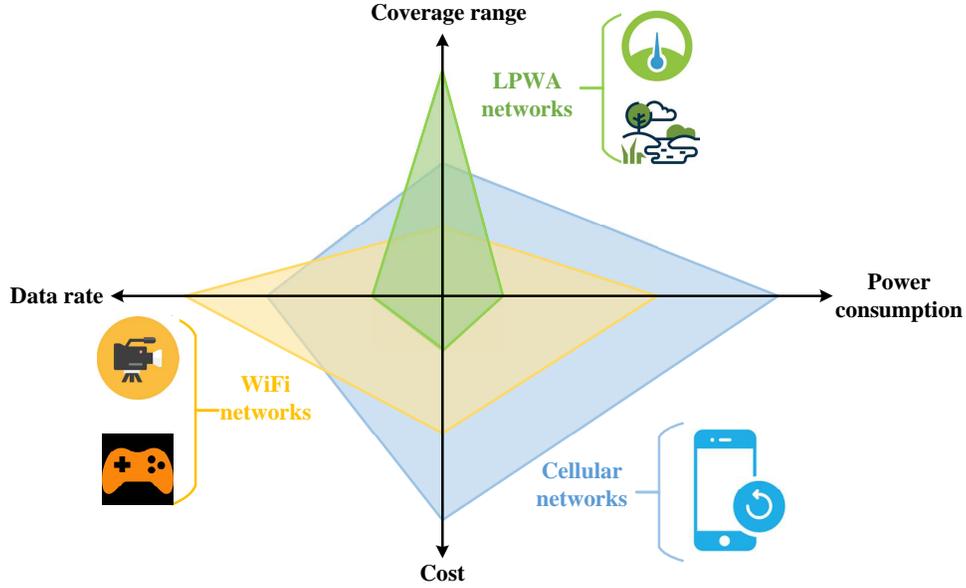}
        \caption{Performance comparison of different wireless networks.}
        \label{adaptive_radio}
    \end{center}
\end{figure}

Various IoT requirements and technologies have resulted in different network designs. For example, IoT devices in smart agriculture and smart city, typically powered by battery, require transmission ranges up to several tens of kilometres and are expected to last for more than 10 years~\cite{3gpp22861}. Therefore, sustainable IoT is more than desired. The legacy cellular network can provide long range coverage, but cannot offer high energy efficiency (EE) due to its complex modulation and multiple access techniques~\cite{Ge:2017}. While typical wireless  local area networks (WLAN), i.e., WiFi, are with low cost, but have very limited coverage. Low-power wide-area (LPWA) technologies are becoming a promising alternative to support sustainable IoT due to its capability to offer better trade-offs among power consumption, coverage, data rate, and cost~\cite{Xiong:2015}. Fig.~\ref{adaptive_radio} compares the performance of different wireless networks from the aforementioned perspectives. As shown in the figure, different networks are developed for different applications. Consequently,  LPWA technologies have the potential to make a significant contribution to IoT ecosystems due to its unique features of  long transmission range and low power consumption.

In LPWA networks, transmission range and power consumption are the two dedicating factors  for highly scalable applications, such as smart monitoring infrastructures, where only a small portion of data is transmitted. In order to address these two issues, two technical possibilities have been proposed. The first approach is to use  the ultra narrrowband technique that enhances  signal-to-noise (SNR) ratio by focusing signal in a narrowband. Narrowband-IoT (NB-IoT)~\cite{3gpp45820} is such an implementation example. Another approach is to utilize coding gain to combat high noise power in a wideband receiver. Long Range  (LoRa)~\cite{LoRa} technique is an example of this that increases transmission range with enhanced power efficiency.

The rest of this article is organized as follows. Section~\ref{s2} overviews the  LPWA candidates that can be implemented for sustainable IoT networks in terms of their fundamental techniques and practical potential. Section~\ref{s3} presents two  implementations of LPWA technology and discussed their usage in large-scale deployment. Section~\ref{s4} concludes this article by identifying several research challenges that remain in the practical deployment of LPWA networks.

\section{Candidates of Low-Power Wide-Area Networks}\label{s2}
In this section, we will compare several promising LPWA candidates that can be implemented to support sustainable IoT networks, including long-term evolution (LTE) machine-type-communications (LTE-MTC/LTE-M), NB-IoT, LoRa, and Bluetooth low energy (BLE). Both the technical basis and their practical deployment potential are discussed and the performance of different radios are compared.

\subsection{Fundamental Techniques and Practical Deployment Potential}
\subsubsection{\textbf{LTE-MTC (LTE-M)}}
LTE-MTC, also called LTE-M, is an  LPWA technology standard published by the 3rd Generation Partnership Project (3GPP)~\cite{3gpp45820}. It works over the licensed spectrum and can achieve a much higher data rate than the other LPWA candidates, such as LoRa, it is therefore  able to transmit heavier data such as video.  What makes LTE-M  an energy efficient LPWA technology is its ability to remain virtually ``attached'' to the network without physically maintaining a connection. Specifically, LTE-M devices can go idle without carrying out the typically heavy processes to rejoin the network on waking-up, and  base station is informed by the endpoint  how often the LTE-M device will be awake. As a result, LTE-M is ideal for mobile devices as it remains the handover mechanism to guarantee seamless coverage.

From the perspective of large-scale implementation, LTE-M for machine-to-machine (M2M) communications  works within the normal infrastructure of LTE networks, which motivates cellular operators, such as Verizon and Vodafone. It only needs to upload new software onto existing base stations without any costs on new infrastructure. Moreover, LTE-M receivers only need to process 1.4 MHz of the channel instead of 20 MHz, which makes it become much simpler than mobile phone receivers.

\subsubsection{\textbf{NB-IoT}}
As an evolution from regular MTC to massive MTC (mMTC), NB-IoT is a 3GPP standard, which was frozen in 2016~\cite{3gpp45820}. It operates over the licensed spectrum, which are allocated for global system for mobile communications (GSM) and LTE.  Therefore, there are two types of NB-IoT technologies:

\begin{itemize}
  \item GSM based NB-IoT: Given that the bandwidth of a GSM channel is 200 kHz, it is straightforward to support NB-IoT through an \emph{in-GSM band} simply by software update on base station. For multiple access schemes, traditional time-division multiple access  and frequency-division multiple access  are employed. For modulation, it adopts Gaussian minimum shift keying (GSMK) or 8 phase-shift keying (8PSK) for both uplink and downlink.
  \item LTE based NB-IoT: In LTE networks, NB-IoT has two operation modes, which can either utilize the unused resource blocks (RBs) within an LTE carrier's guard-band or part of an LTE carrier with a self-contained NB-IoT cell. For downlink transmission, NB-IoT adopts quadrature phase-shift keying (QPSK) modulation. For uplink transmission, either binary phase-shift keying (BPSK) or QPSK modulation is adopted. When multiple IoT devices request LTE connection for link establishment and resource allocation, a random access procedure is used. However, the random access mechanisms cannot handle simultaneous connections for massive IoT as only a limited number of preambles are available for MTC. With a large number of simultaneous access attempts, high probability of collisions may occur~\cite{survey14}.
\end{itemize}

The high compatibility with GSM  and LTE  provides NB-IoT  with great potential for  worldwide deployment. In some countries, such as Norway, where GSM networks have been phased out, both LTE-M and NB-IoT services can be provided purely based  on 4G networks. In other countries where 2G or/and 3G are still operational, NB-IoT services can be supplied based on these existing infrastructures.

\subsubsection{\textbf{LoRa}}
 Different from LTE-M and NB-IoT, LoRa works over an unlicensed spectrum, i.e., 433~MHz, 868~MHz, and 915~MHz. As defined by LoRa Alliance in 2015~\cite{LoRa}, the chirp spread spectrum technology  is used in LoRa, which allows the usage of cheap oscillators but with high stability  at the receiver. As a result, the cost of each LoRa device is lowered. Additionally, different users can share the same RB by using different spreading factors (SFs). Moreover, different SFs require different sensitivities at the receiver, which leads to  different transmission ranges and data rates. Note that LoRa is more robust to burst noise than NB-IoT as  LoRa channels span over a wider bandwidth.

\begin{figure}[t!]
    \begin{center}
        \includegraphics[width=3.6in,height=2.5in]{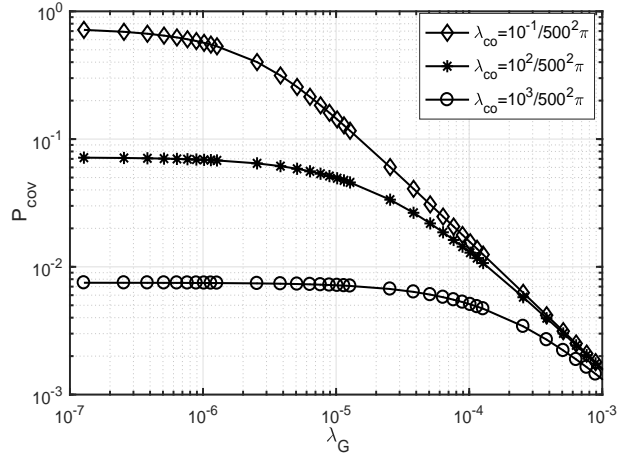}
        \caption{Coverage probability $P_{cov}$ versus density of LoRa devices $\lambda_G$ with different densities  of non-LoRa devices $\lambda_{\rm co}$, which are working over the same frequency. In the figure, LoRa device transmit power is 14~dBm, threshold is 20~dB, channel bandwidth is 125~kHz with frequency 868~MHz, and the LoRa devices distribution range is up to 20~km~\cite{Zhijin:ICC2017}}.
        \label{interference}
    \end{center}
\end{figure}

Due to the high flexibility and scalability, LoRa has attracted increasing interests from  both industry and academia, making it as one of the most widely deployed LPWA technologies around the world. However, interference from LoRa devices sharing the same time/frequency slot and same/different SFs will deprave  channel environment and lower the coverage probability\footnote{The coverage probability is defined as the probability that the transmit signal of a typical LoRa user can be successfully decoded at the gateway. In another word, the SINR of the LoRa user received at the gateway is higher than the predefined threshold.} of LoRa gateways.  Fig.~\ref{interference},  from~\cite{Zhijin:ICC2017}, shows that the coverage probability  of a LoRa gateway  drops significantly with the increasing density of LoRa devices and that of non-LoRa devices transmitting over the same frequency. As the densities of LoRa and non-LoRa devices increase, both the intra and inter interference become heavier. The usage of the unlicensed spectrum makes the interference issue more critical in LoRa networks, especially when non-LoRa devices also work over the same frequency simultaneously.  In order to provide more insights for the implementation of LPWA technologies, we will provide a case study of LoRa  in Section~\ref{lora-reality}.

\subsubsection{\textbf{BLE}}
BLE was released in 2010 as part of the Bluetooth 4.0 core specification \cite{bluetooth4}, but it is not backwards compatible with the classic Bluetooth. BLE works over the unlicensed 2.4 GHz band and the transmission range can be up to 100~m. With Bluetooth 5 released in 2016, the transmission range could be quadrupled through increased transmit power or coded physical layer. In 2017, the Bluetooth special interest group  ratified a mesh specification, allowing multi-hop communication based on BLE technologies. 
In comparison with the classic Bluetooth, BLE is characterized as ultra-low peak, average, and idle current consumption, allowing BLE devices to run for years on standard coin-cell batteries.

BLE has  various applications with cost-low and battery-powered devices, e.g. healthcare, remote control, and smart home. A unique feature of BLE, in contrast to LoRa, is that a majority of mobile phones already support BLE. With its existing cellular connection, a mobile phone can naturally act as a relay node for providing LTE connection to indirect 3GPP IoT devices. An example will be given in Section \ref{wur}.

\newcommand{\tabincell}[2]{\begin{tabular}{@{}#1@{}}#2\end{tabular}}
\begin{table}[t!]
\centering
\caption{Comparison of different LPWA candidates}
\label{Comparison}
\begin{tabular}{|l|l|l|l|l|}
\hline
LPWA technology &  LTE-M  &NB-IoT  &  LoRa&  BLE    \\ \hline
Spectrum    &  \tabincell{l}{Licensed\\ LTE in-bands}  & \tabincell{l}{Licensed\\LTE in-bands, guardband\\ GSM standalone} & \tabincell{l}{ Unlicensed\\ 433/868/915 MHz} &\tabincell{l}{ Unlicensed\\2.4 GHz}  \\ \hline
    Bandwidth    &      1.4 MHz  & 180 kHz   &  125/250/500~kHz  &      2 MHz               \\ \hline
      Max output power  &   20 dBm &   23 dBm   &  14 dBm  &      10 dBm              \\ \hline
      Date  rate       &   1 Mbps &   0.3$\sim$50 kbps   &  up to 250 kbps &      125 kbps/1 Mbps/2 Mbps  \\ \hline
      Link budget    &     146 dB  &  	164 dB  &  154 dB  &    105 dB      \\ \hline
      Power efficiency     &     Medium  &   Very high   &  Very high  &     Medium        \\ \hline
     Interference immunity &     Low  &   Low   &  Very high  &    Very high       \\ \hline
     Security &    Very high  &   Very high  &  Low &    Low       \\ \hline
     Network cost &    Low  &   Low &  High &    High       \\ \hline
     Standardization&     3GPP Release 13   &   3GPP Release 13   &  LoRA Alliance  &     Bluetooth 4.0     \\ \hline
\end{tabular}
\end{table}

\subsection{Performance Comparison}
Due to the increasing market for LPWA networks, intense competition among network operators and new LPWA network offerings has created more cost-effective technologies. Table~\ref{Comparison} compares different characteristics of the aforementioned LPWA technologies. Note that they can achieve different data rates and coverage ranges, which makes them ideal for different application scenarios. As supported by different standard organizations and alliances, each LPWA technology shows different practical deployment potential. For example, LTE-M and NB-IoT maybe favoured by cellular network operators due to their high compatibility with existing cellular networks. While LoRa and BLE are supported by IoT operators with specific applications due to their good network scalability. So far, much of Europe and many parts of the Asia Pacific region are covered by LoRa networks with LPWA services. While in the U.S.A., LoRa will have the largest private unlicensed industrial LPWA network growth and LTE-M will do the same for the licensed public networks.

\section{Low-Power Wide-Area Technologies in Reality}\label{s3}

This section provides two case studies of LPWA in reality, including LoRa for smart city and wake-up radio critical infrastructure surveillance. Then we discuss the insights obtained from the implementation of LPWA networks.

\subsection{LoRa Reality for Smart City}\label{lora-reality}
For green networks with power-constrained  IoT devices, one of the design paradigms is to maximize the battery lifetime. Smart resource allocation, i.e., duty cycle control and user scheduling, is an approach to achieve this. Additionally, the recent developments on energy harvesting technologies, including radio frequency based wireless charging, provide another way to further extend lifetime of wireless sensors by introducing extra energy.

In order to realize sustainable IoT networks, we have introduced energy neutral operation (ENO) to the field test in the living lab established in the Queen Elizabeth Olympic Park (QEOP), London. ENO enables the continuous operation of wireless sensor networks. However,  note that the battery capacity degrades over time. Rather than optimizing the operation within one period, we have been focused on maximizing the quality-of-service (QoS) in terms of duty cycle and the battery capacity over the lifetime of a sensing application. A lightweight algorithm has been deployed to extend lifetime of the various sensing applications. Our test results in~\cite{ICDCS:2017} have shown that a significant extension of deployment lifetime can be achieved without a reduction in the duty cycle of the sensor.

\begin{figure}[!t]
    \centering
    \subfigure[LoRa-aided data collection framework.]{
    \includegraphics[width= 2.7in]{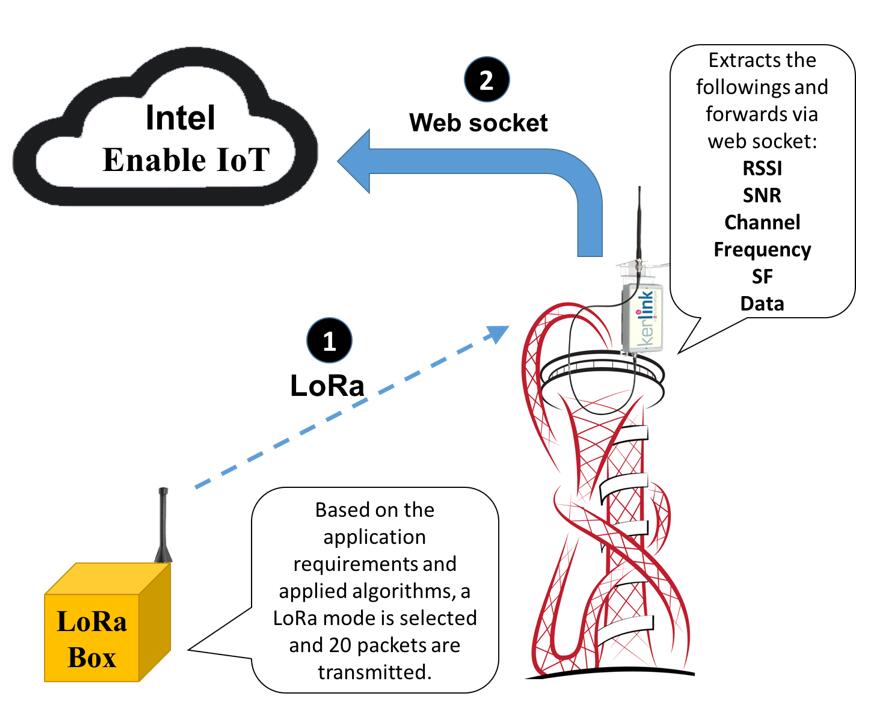}}
    \subfigure[Packet transmission over channel 14 at 868 MHz with bandwidth 500 kHz, SF=7, CR=4/5, transmit power is 0 dBm.]{
\includegraphics[width= 3.6in]{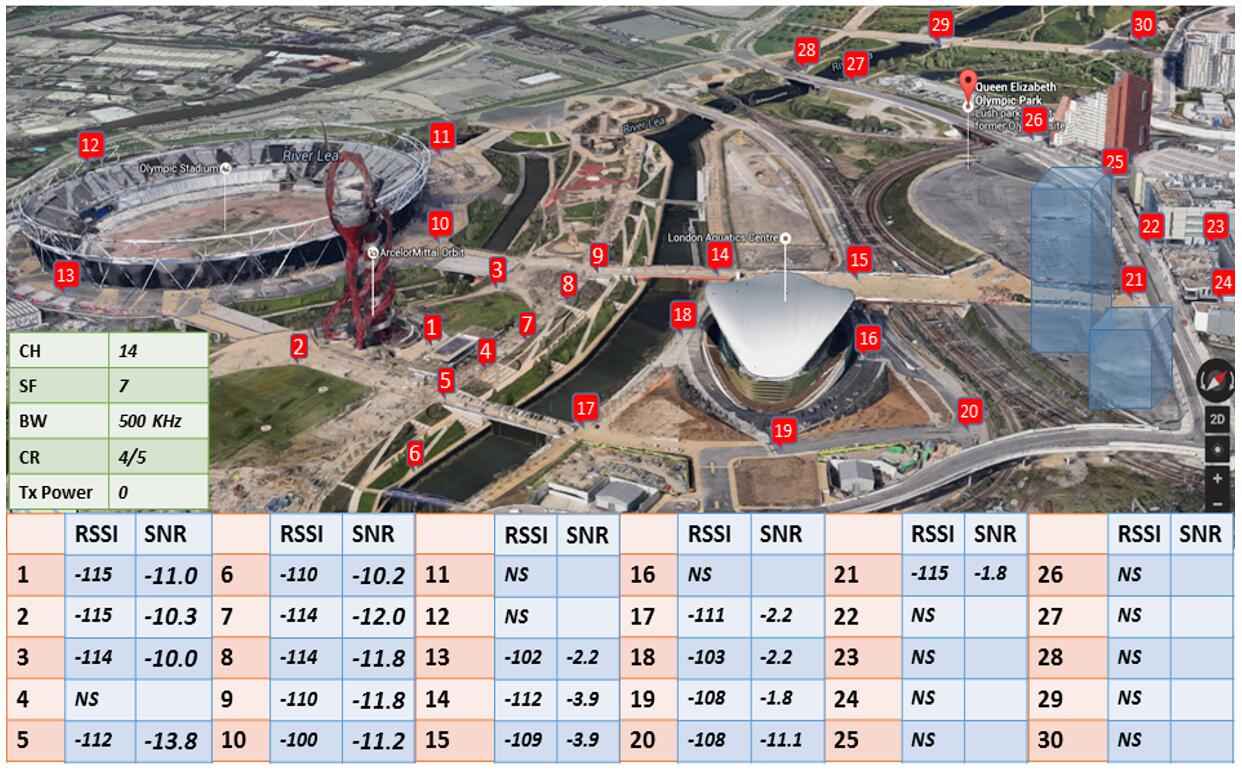}}
\caption{Deployment of LoRa devices in the Queen Elizabeth Olympic Park, London~\cite{Mo:2017}.}
\label{LoRa_box}
\end{figure}

In addition to the ENO enabled battery lifetime extension, how to adaptively schedule sensor transmissions to improve network scalability becomes another important issue. As shown in  Fig.~\ref{LoRa_box} (a), a LoRa gateway has been installed on top of the Orbit  and 30 LoRa devices, named LoRaBox, have been deployed in the QEOP for data collection. Based on the application requirements, the LoRaBox selects a proper mode, e.g., bandwidth, coding rate, and SF, to upload the collected data to the gateway. Then the gateway forwards the received data to the cloud server for further processing and analyzing~\cite{Mo:2017}. Fig.~\ref{LoRa_box}(b) shows the deployment locations of LoRaBox in the park and the RSSI and SNR values of these LoRa nodes received at the gateway. In this case, it is noted that signals from LoRaBox experiencing poor channel conditions, i.e., long transmission distance, deep fading and shadowing, cannot reach the LoRa gateway. For example, `LoRaBox 30' is too far from the gateway and `LoRaBox 12' is blocked by the buildings. In order to improve the transmission quality, the channel, SF, and transmit power should be carefully selected. For instance, a higher SF, i.e., increasing from $7$ to $12$, and higher transmit power, i.e., increasing from 0~dBm to 14~dBm, should be used to extend the transmission range.

\begin{figure}[!t]
    \centering
    \subfigure[Performance of the proposed low-complexity channel assignment scheme based on matching theory.]{\label{fig.a}
    \includegraphics[width= 3.1in]{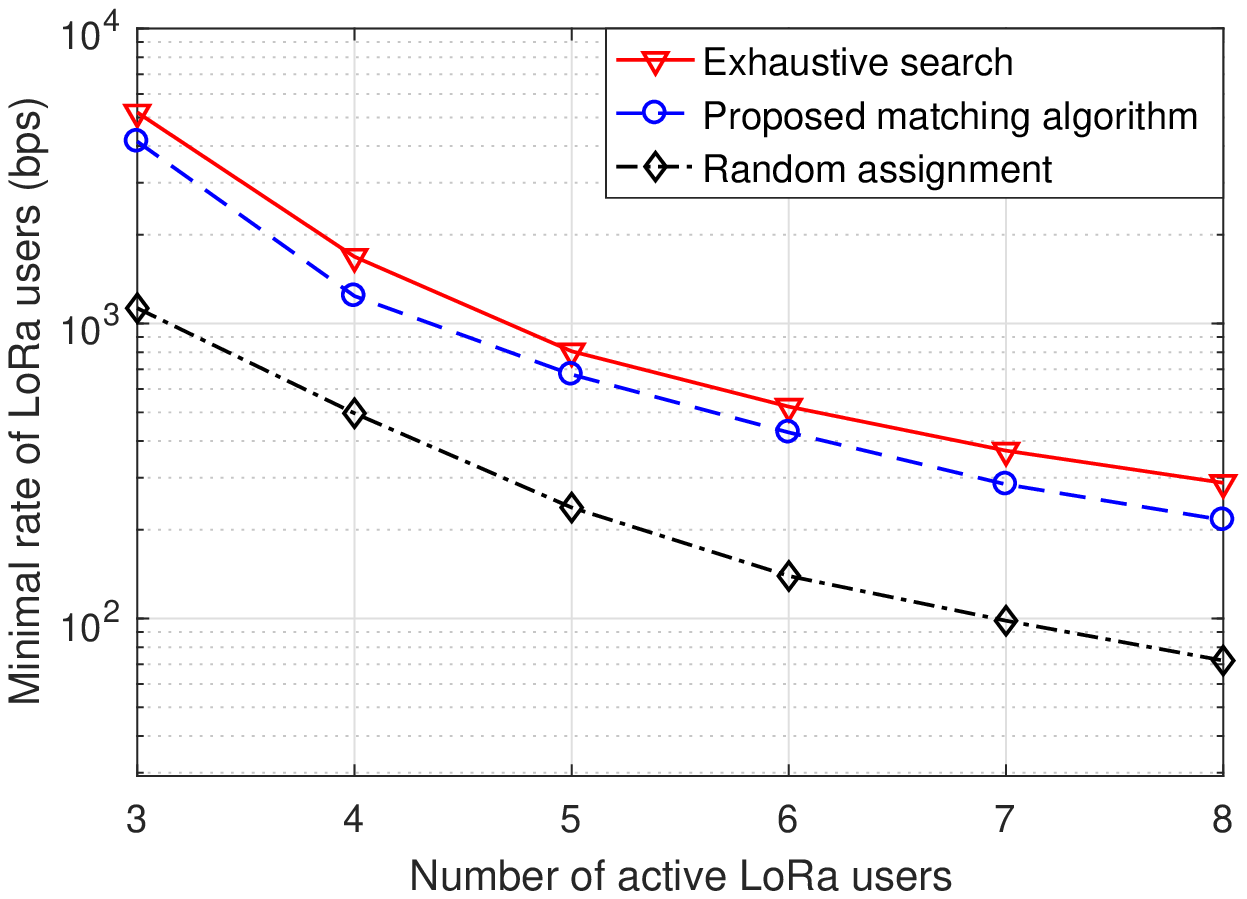}}
    \subfigure[Performance of proposed optimal power allocation scheme for users assigned to the same channel.]{\label{fig.b}
\includegraphics[width= 3.1in]{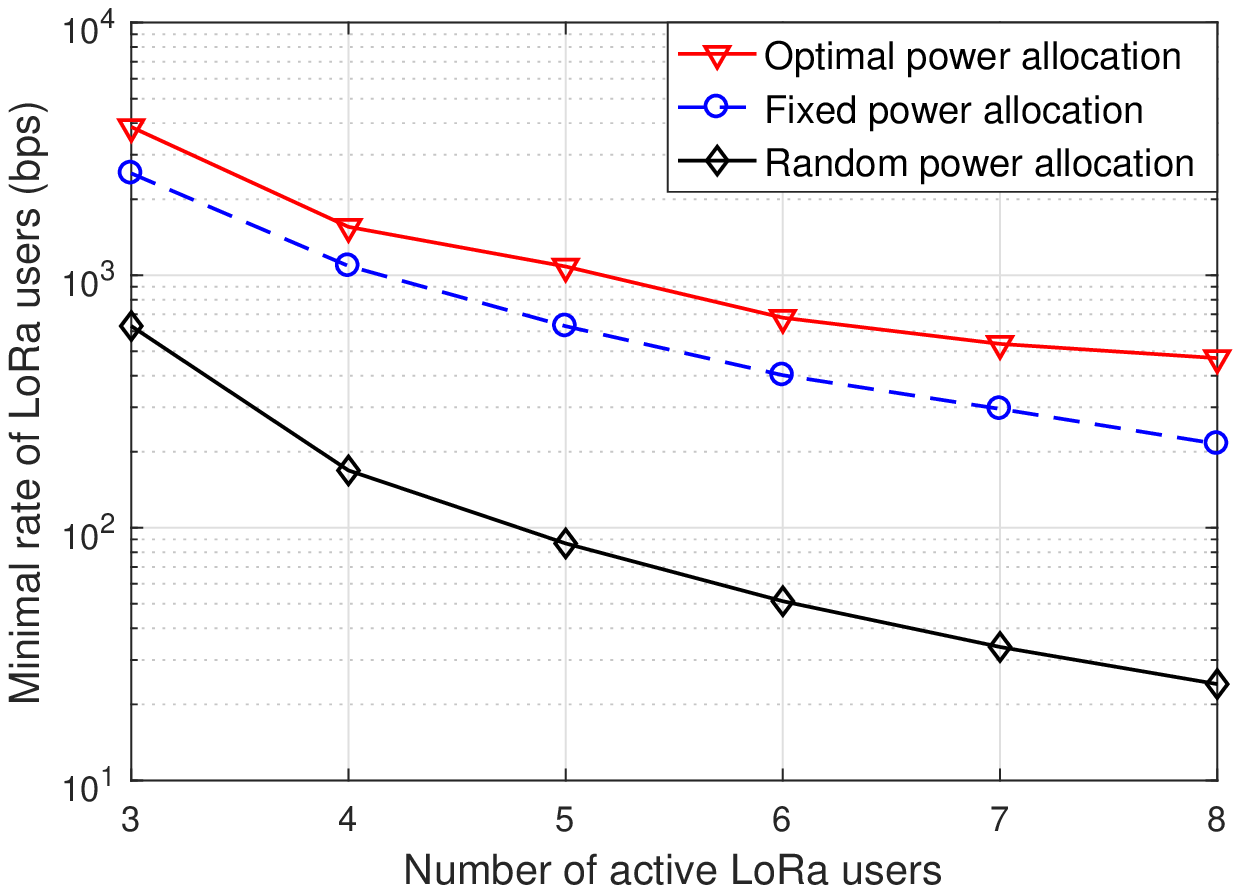}}
\caption{Active LoRa nodes are randomly distributed in a circle with the radius of 10 km, center frequency  is 868 MHz, the number of available channels is 3 with bandwidth of 125 kHz for each, the transmit power of each LoRa users ranges from 0~dBm to 20~dBm.}
\label{RE}
\end{figure}

In order to manage LoRa devices efficiently,  a low-complexity algorithm has been proposed to guarantee the transmission rate fairness among LoRa devices in \cite{Zhijin:GC2017}. Particularly, the user fairness problem in LoRa networks is formulated as a joint problem of channel assignment and power allocation. With the purpose of maximizing the minimum data rates of LoRa users, each LoRaBox configures itself with the proper channel by considering its preference and the other nodes' choices. Matching theory has been invoked to realize the channel assignment problem. From Fig.~\ref{fig.a}, the proposed channel assignment scheme can achieve near-optimal performance in comparison with the exhaustive search approach while the required complexity is much lower as shown in \cite{Zhijin:GC2017}.  Once LoRa devices are assigned with proper channels, their transmit powers  are then optimized to improve the system performance. As shown in Fig.~\ref{fig.b}, the developed optimal power allocation scheme outperforms that with the fixed and random power allocation, respectively.

\subsection{Wake-up Radio for Critical Infrastructure Surveillance}
\label{wur}
Duty cycle medium access control mechanisms, which allow nodes sleep and wake up cyclically, have been a legacy solution  to reduce energy consumption to support sustainable IoT networks, e.g., in the first example given above. However, such mechanisms suffer from idle listening and overhearing, as well as long delay since packets can only be transmitted when nodes are awake. Recently a paradigm shift from duty cycled mechanisms to wake-up radio (WuR) has been envisaged \cite{oller2016has}. Thanks to WuR's convincing superiority for energy consumption, which is in the magnitude of 1,000 times lower than that of the traditional radios.

Consider a scenario where a large volume of WuR-enabled IoT devices are deployed to monitor the operation of a critical infrastructure, e.g., a tunnel or a bridge. The IoT devices do not support direct 3GPP connection due to the operation cost for maintaining cellular connections and power consumption cost for uplink traffic. Nor is it necessary to transmit data at a frequent interval considering the sporadic feature of traffic intensity for such IoT applications. To gather data in such a network, one may perform either transmitter-initiated data reporting or receiver-initiated data collection.

\begin{figure}[!t]
    \centering
	\includegraphics[width=3.8in]{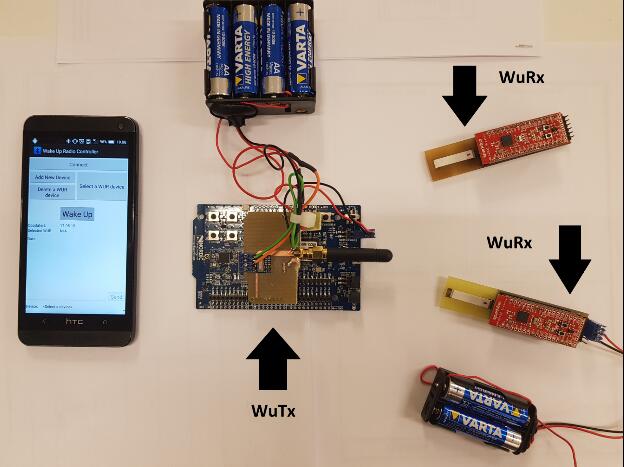}
	\caption{Illustration of a WuR-enabled IoT testbed with indirect 3GPP connection where the mobile phone serves as a relay for cellular connection.}
	\label{fig:prototype}
\end{figure}

Figure \ref{fig:prototype} illustrates a real-life prototype WuR-based test-bed at the University of Agder. The test-bed consists of a smart phone with a BLE connection to the WuR transmitter (WuTx) and multiple WuR receivers (WuRxs). While the main radio for data transmission is kept asleep, the WuRxs are always active with an ultra-low power consumption level of 1.3 $\mu$W. This means that the lifetime of these devices could be much longer that 10 years as required for massive IoT~\cite{3gpp22861}. The mobile phones function as a relay between LTE and these non-3GPP IoT devices. In the considered network, a receiver-initiated data collection procedure has been designed. Particularly, a~mobile data collector carries a mobile phone and requests one or multiple devices to report their stored data via a wake-up call (WuC). Such a mobile data collector could be for instance a vehicle or a drone. Once the  WuC is received, the WuRx will trigger its main radio to perform data communication towards the relay. To avoid collision when multiple IoT devices wake up and attempt to transmit at the same time, a backoff mechanism is desired~\cite{tii18}. Depending on the traffic load status, we have proposed three protocols which employ either clear channel assessment (CCA) alone, CCA together with backoff, or CCA with an adaptive backoff procedure, to avoid the WuC collision~\cite{tii18}.

\subsection{Insights from IoT Projects}
In addition to the aforementioned examples, there have been many undergoing IoT projects targeted at different applications, such as smart agriculture, disaster detection, and digital  health, which are funded by both industrial and academic organizations. The aforementioned implementation example of LoRa is a five-year project on smart city with industry and academic collaborations\footnote{More details of the project can be found at http://cities.io/}. Various projects with similar purposes have been supported by different companies, such as Intel and IBM, and they have been carried out in different cities around the world to show the potential of deploying LPWA networks in a large scale. Another well-known example of IoT project on smart agriculture is that Fujitsu has launched a company in Finland to produce and sell vegetables year-round with artificial-light plant factory\footnote{More details can be found at http://www.fujitsu.com/global/about/resources/news/press-releases/2016/1128-01.html}. In order to make up the shortage of leaf vegetables in northern, sensors have been deployed in the greenhouse to collect information and a cloud server has been built up to realize automatic control of environment in the greenhouse. All these IoT projects give us the confidence to deploy LPWA networks in the real world.

In the meantime,some IoT projects experienced failure  due to the dramatic difference between theory and practice. Moreover, the various implementation environments and hazards are also common reasons for failures. For example, the sensors equipped with energy harvesting component was covered by a plastic shell for water-proof purpose, which the water-proof shield may lead to inaccurate sensor readings in terms of the temperature and humility in the wild. So it is very important to consider various factors in deployment. Failures may come from improper deployment process even though advanced algorithms have been developed to magnae sensors behaviours.

\section{Research Challenges and Conclusions}\label{Challenges}\label{s4}
Before  large-scale LPWA technologies are deployed in the real world, the following research challenges should be  addressed properly.
\begin{enumerate}
  \item \textbf{Spectrum  and EE optimization}: As LPWA networks aim to serve massive number of battery-powered IoT devices by limited spectrum resources, spectrum efficiency (SE) and EE become two  critical issues\footnote{Please refer to \cite{CHEN:CM:2011} to get better understanding of the relationship between SE and EE.}. Therefore, optimizing SE and EE by considering the joint design of physical layer transmission and resource management become essential in LPWA networks. Two potential research directions are summarized in the following where LoRa is considered as the focus.
\begin{itemize}
  \item \textbf{Interference management}: In order to optimize the SE and EE,  interference from LoRa devices using the same SF and different SFs served by the same gateway should be considered. This can be formulated as a joint problem to optimize the user association and transmit power. As a result, the designed LoRa network should have the capability to: i) guarantee the required QoS, i.e., different SNR requirements for different SFs, for each user; ii) optimize transmit power of each user to offer massive connectivity; iii) reduce the SF conflict probability to minimize inventable interferences.
  \item \textbf{User scheduling}: Transmission conflict is a challenging issue in LPWA networks where there are a massive number of LoT devices attempting to transmit data even though the packet is usually small. The user scheduling problem can be formulated as a channel allocation, SF assignment, and transmit power optimization problem. With the proper design, LPWA network should be able to: i) assign the most suitable channel to each LoRa device; ii) fully utilize the transmission features of different SFs. For instance, devices with a short distance to the gateway may prefer a lower SF; iii) optimize transmit power for each user with guarantee on the required QoS.
\end{itemize}
Furthermore, compared with the traditional optimization tools, machine learning can be potentially used for resource management for LPWA networks as the high complexity training process can be carried out offline and each IoT device only needs to use the trained model for decision making.
\item \textbf{Adaptive LPWA radio selection}: Rather than using any single LPWA technology, multi-radio modules are creating new opportunities for industrial IoT developers. The first variation can be multi-radio LPWA modules with integrated short-range radios, such as BLE and 802.15.4. Another trend of LPWA hybrid modules is to use two LPWA radios, including a low-bit-rate radio for long device lifetime functions and a high-bit-rate low-latency radio for advanced functions and over-the-air updates.

  \item \textbf{Distribution and implementation}: Distributed frameworks are highly desired to enable the implementation in ultra-dense networks for IoT applications. In-field implementations of the LPWA networks, i.e., NB-IoT networks, with  distributed resource management algorithms, are of great interest and are expected to bring the research from theory to practice.
\item \textbf{Sensing without sensors}: Thanks to the advanced functions of mobile phones, they can be adopted as mobile sensors by considering the mobility of the human carrier. Therefore, an opportunistic network becomes available for data collection. By properly designing crowdsourcing mechanisms, we can save some IoT sensor devices in certain scenarios.
\end{enumerate}

In summary, this article has provided an overview on the existing solutions to LPWA networks. Moreover, two case studies have demonstrated the potential of implementing LPWA technologies for real-life applications. Based on the analysis on technical fundamentals  and the practical deployment of LPWA technologies, we have identified several research challenges to be addressed before bringing LPWA networks from theory to practice.

\bibliographystyle{IEEEtran}

 \end{document}